%% file: main.tex
\documentclass[11pt,a4paper]{article}

\usepackage[T1]{fontenc}
\usepackage{amsmath,amssymb,amsfonts}
\usepackage{bm}
\usepackage{booktabs}
\usepackage{graphicx}
\usepackage{xurl}
\usepackage[colorlinks=true,linkcolor=blue,citecolor=blue,urlcolor=blue]{hyperref}
\usepackage[authoryear,round,comma,sort&compress]{natbib}
\usepackage{geometry}
\usepackage{setspace}
\usepackage{caption}
\usepackage{subcaption}
\usepackage{enumitem}
\usepackage{bbm}
\usepackage{xspace}
\usepackage{multirow}
\usepackage{float}

\title{\textbf{Cross-Bond Panel Learning for Next-Day Price Forecasting\\of Chinese Convertible Bonds: A Reality Check}}

\author{
  Derun Zhong$^{1\dagger}$, Guolian Wang$^{1*}$, Shiqing Wang$^{2}$, and Hongjie Feng$^{1}$\\[6pt]
  $^{1}$School of Mathematical Sciences, Tongji University, Shanghai, China.\\
  $^{2}$School of Economics and Management, Tongji University, Shanghai, China.
}

\date{}

\begin{document}

\maketitle

\begingroup
\renewcommand\thefootnote{}%
\footnotetext{$^{\dagger}$Derun Zhong, Email: \href{mailto:1928027160@qq.com}{1928027160@qq.com}.}%
\footnotetext{$^{*}$Guolian Wang, Corresponding author, Email: \href{mailto:wgl@tongji.edu.cn}{wgl@tongji.edu.cn}.}%
\footnotetext{This work was supported by the National Undergraduate Innovation and Entrepreneurship Training Program of Tongji University in 2025 (Project No. 202510247114).}%
\endgroup

\begin{abstract}
Recent advances in financial machine learning have prompted growing interest in applying deep neural networks to convertible bond valuation and forecasting. However, existing studies often focus on in-sample valuation fit or cross-sectional return rankings, leaving the foundational question of out-of-sample next-day price forecasting insufficiently benchmarked against persistence. This paper provides a comprehensive empirical reality check on cross-bond panel learning for Chinese convertible bonds. Using a daily panel of 73,806 observations across 100 A-share convertible bonds from 2022 to 2025, we evaluate whether cross-bond neural architectures can reliably outperform the random walk under strict temporal splitting, boundary purging, fold-specific preprocessing, and dependence-aware inference. Across the 29,514-observation holdout period, the random walk achieves an RMSE of 2.3577 CNY. Five-seed ensembles of small and large price-change MLPs yield RMSEs of 2.3619 and 2.3645 CNY, corresponding to negative out-of-sample $R^2$ values of $-0.0035$ and $-0.0057$ and target-date clustered HAC $p$-values of 0.453 and 0.248. Ten-day circular block-bootstrap intervals span zero, while direct price-level training degrades performance substantially. Although cross-bond pooling provides a variance-stabilizing shrinkage effect over isolated single-bond estimation—lowering test RMSE for 79 of 90 individual bonds—it fails to produce an aggregate forecast advantage over persistence. Error decomposition reveals that failure is not uniform: five highly volatile, clause-sensitive bonds account for 74.4\% of the model's total excess squared error. These findings indicate that without contemporaneous equity market information or explicit contractual event conditioning, deep panel architectures do not deliver incremental next-day predictability over persistence, establishing a rigorous benchmark for future quantitative convertible bond research.

\vspace{6pt}
\end{abstract}

\input{sections/01_introduction}
\input{sections/02_related_work}
\input{sections/03_methodology}
\input{sections/04_experimental_setup}
\input{sections/05_results}
\input{sections/06_discussion}
\input{sections/07_conclusion}
\input{sections/08_appendix}

\section*{Acknowledgements}
This work was supported by the National Undergraduate Innovation and Entrepreneurship Training Program of Tongji University in 2025 (Project No. 202510247114).

\section*{Disclosure of Interest}
The authors report there are no competing interests to declare.

\bibliographystyle{plainnat}
\bibliography{references}

\end{document}

%% file: sections/01_introduction.tex
\section{Introduction}
\label{sec:intro}

Convertible bonds (CBs) are hybrid contingent claims that combine fixed-income downside protection with equity upside participation through embedded conversion options. In China's rapidly expanding exchange-traded market, convertible bonds feature distinct contractual mechanisms, including issuer call provisions, investor put options, and issuer-initiated downward conversion price resets. This intricate path dependence and game-theoretic complexity have spurred substantial interest in machine-learning and deep-learning approaches for pricing and predictive modeling~\citep{tan2022,ren2023,zhu2024,li2026gan}.

A growing literature asserts that deep neural architectures can capture complex nonlinearities in convertible bond markets, frequently reporting high directional accuracy or close valuation fit~\citep{tan2022,ren2023,zhu2024}. However, existing empirical work predominantly focuses on contemporaneous valuation fitting against theoretical models or weekly cross-sectional return rankings~\citep{li2026returns}. The foundational question of out-of-sample next-day price forecasting—whether machine-learning panel models can reliably beat the simple persistence forecast $P_{t+1}=P_t$ (the random walk)—remains systematically unaddressed. In financial economics, traded prices over daily horizons follow near-martingale dynamics~\citep{fama1970,meese1983}. Outperforming persistence requires genuine incremental predictive signal rather than the mere recreation of a persistent price level. Furthermore, subtle methodological vulnerabilities, such as look-ahead bias from same-day covariates, information leakage across temporal split boundaries, or global feature preprocessing, can easily inflate apparent forecast performance.

This study provides an exhaustive, leak-resistant empirical evaluation of cross-bond panel learning for next-day price forecasting in the Chinese convertible bond market. Specifically, we investigate whether pooling panel data across multiple convertible bonds enables neural regressors to extract stable, generalizable pricing signals that outperform the random walk benchmark once temporal boundaries, feature scaling, network capacity, and cross-sectional dependence are rigorously controlled.

Our empirical architecture incorporates several essential econometric safeguards:
\begin{enumerate}[leftmargin=*,itemsep=2pt]
\item \textbf{Strict temporal hygiene and boundary purging:} Forecasts are anchored to explicit information dates $t$ and target dates $t+1$. Observations spanning temporal split boundaries are purged, and all preprocessing scalers are refit exclusively on historical training data for every window and fold.
\item \textbf{Econometrically grounded benchmarks:} Cross-bond multi-layer perceptrons (MLPs) are systematically compared against persistence (the random walk), historical drift, pooled AR(1), and regularized linear baselines (Ridge and Elastic Net).
\item \textbf{Direct target ablation:} We contrast models trained to predict residual price changes ($\Delta P_{t+1} = P_{t+1} - P_t$) against models trained directly on next-day price levels ($P_{t+1}$), highlighting the severe estimation penalty incurred when neural networks attempt to learn an identity mapping.
\item \textbf{Controlled pooling comparisons:} To isolate the specific contribution of cross-asset panel learning from differences in network capacity or optimization, we benchmark pooled models against capacity-matched, bond-specific estimators under paired observation sets and fixed optimizer-update budgets.
\item \textbf{Dependence-aware statistical inference:} To account for cross-sectional market shocks that induce correlation across bonds on identical target dates, we employ target-date clustered Newey--West (HAC) tests weighted by daily cross-sectional frequency, complemented by circular and moving block bootstraps.
\end{enumerate}

Our findings establish that cross-bond neural networks do not provide reliable out-of-sample predictive gains over persistence. Across a 29,514-observation holdout panel spanning July 2024 through December 2025, the random walk achieves an RMSE of 2.3577 CNY. Five-seed ensembles of small and large price-change MLPs produce RMSEs of 2.3619 CNY ($R^2_{\mathrm{OOS}} = -0.0035$) and 2.3645 CNY ($R^2_{\mathrm{OOS}} = -0.0057$), with dependence-aware HAC $p$-values of 0.453 and 0.248. Ten-day circular block-bootstrap intervals span zero, and one-sided 95\% upper bounds constrain any plausible MSE improvement to less than 0.45\%. Direct price-level training performs substantially worse, with ensemble $R^2_{\mathrm{OOS}}$ dropping to $-0.1042$ and deteriorating further under high-dimensional feature spaces. Walk-forward evaluations reveal that brief periods of apparent outperformance (such as 2024H2) fail to persist across subsequent market regimes.

Importantly, our analysis reveals two crucial structural insights into financial panel learning. First, cross-bond pooling exerts a clear variance-stabilizing shrinkage effect: estimating a shared architecture across the panel lowers test RMSE for 79 out of 90 individual bonds compared to estimating separate networks per bond. Second, forecast failure is heavily concentrated in the tails: just five highly volatile, clause-sensitive convertible bonds account for 74.4\% of the total excess squared error generated by the pooled neural model. These instruments experienced speculative trading surges into high-convexity regimes with daily price jumps exceeding 25 to 50 CNY. Without explicit point-in-time modeling of contractual events (such as forced redemption calls or conversion resets) or contemporaneous equity returns, continuous neural approximations incur severe squared loss on discrete price jumps.

\subsection{Contributions}

This study makes four primary contributions to the literature on financial machine learning and empirical asset pricing:
\begin{enumerate}[leftmargin=*,itemsep=2pt]
\item \textbf{Establishing a rigorous out-of-sample benchmark:} We provide the first leak-resistant empirical evaluation of cross-bond neural forecasting against persistence in the Chinese convertible bond market, demonstrating that high in-sample fit or price-level $R^2$ cannot be equated with out-of-sample forecastability.
\item \textbf{Disentangling pooling shrinkage from market predictability:} We show that cross-sectional panel learning meaningfully stabilizes estimation across noisy individual securities, but this stabilization does not suffice to overcome the near-martingale barrier of daily price formation.
\item \textbf{Uncovering the micro-structure of forecast failure:} Through row-level error decomposition, we identify that neural forecast disadvantages are driven by extreme tail events and speculative clause-driven regimes, rather than uniform model miscalibration.
\item \textbf{Methodological standards for derivative panel forecasting:} We demonstrate the necessity of target-date alignment, boundary purging, fold-local scaling, and target-date clustered HAC inference, providing a reproducible methodological blueprint for empirical derivative studies.
\end{enumerate}

%% file: sections/02_related_work.tex
\section{Related Work}
\label{sec:related}

We organize the literature into four themes: convertible bond valuation and forecasting, financial machine-learning benchmarks, panel learning, and technical indicators. This structure separates studies of theoretical or model-implied value from our narrower task of forecasting the next observed closing price.

\subsection{Convertible Bond Pricing and Forecasting}

Classical contingent-claims models originate with \citet{ingersoll1977} and \citet{brennan1977}. Chinese convertible bonds create additional valuation challenges because issuer calls, investor puts, and reset clauses make payoffs path dependent. Recent studies have used generative return models, LSTM and GAN models, learning-based Monte Carlo methods, a GAN--Transformer--LSM architecture, and a neural PPDE solver to address this valuation problem~\citep{tan2022,ren2023,zhu2024,li2026gan,zhu2026ppde}. These papers compare theoretical or model-implied values with alternative valuation procedures or observed market prices. While these studies advance the computational modeling of complex contractual terms, their primary objective is contemporaneous valuation fitting or numerical simulation, rather than out-of-sample next-day price forecasting relative to persistence. Demonstrating that a neural network can approximate structural pricing formulas in-sample does not imply that it extracts tradeable or incremental directional signals for future closing prices.

\subsection{The Importance of Baselines in Financial ML}

Benchmark choice determines the meaning of a reported improvement. \citet{gu2020} compare machine-learning stock-return forecasts with linear alternatives, and \citet{bucci2020} evaluates neural volatility forecasts against established econometric models. For convertible bonds, \citet{li2026returns} predicts weekly cross-sectional returns with 13 machine-learning methods and assesses both out-of-sample $R^2$ and portfolio performance. That return-ranking task differs from our next-day price-level forecast, for which the natural no-change benchmark is the current close.

The random walk, or more generally the persistence forecast, has been a standard benchmark in exchange-rate forecasting since \citet{meese1983}. Its demanding performance is precisely why a next-day level forecast should report incremental accuracy relative to persistence rather than rely on a high conventional level $R^2$.

\subsection{Panel Data and Multi-Task Learning in Finance}

Panel estimation allows financial models to learn common relations from many instruments, as in cross-sectional asset pricing~\citep{gu2020,li2026returns}. Neural networks have also been used for derivative valuation and volatility forecasting~\citep{ferguson2018,bucci2020}. Pooling may improve prediction relative to separate small-sample fits even when the learned common relation has little value relative to persistence. We therefore distinguish the pooled-versus-single forecast comparison from the separate question of whether either forecast beats the random walk.

\subsection{Technical Indicators in Financial Prediction}

Technical indicators are widely used in financial forecasting. Prominent examples include RSI~\citep{wilder1978}, MACD~\citep{appel1979}, and Bollinger Bands~\citep{bollinger1992}. Recent deep-learning studies have incorporated such indicators into forecasting pipelines~\citep{sezer2020}. Our ablation asks whether they add predictive value beyond price lags for convertible bonds.

%% file: sections/03_methodology.tex
\section{Methodology}
\label{sec:method}

\subsection{Forecasting Problem and Targets}
\label{sec:problem}

Let $P_t^{(i)}$ denote the closing price of bond $i$ on information date $t$. Each forecast is formed immediately after that close and concerns the next date represented in the market-date sequence, denoted $t+1$. The primary target is the next-day price change,
\begin{equation}
\Delta P_{t+1}^{(i)} = P_{t+1}^{(i)}-P_t^{(i)},
\label{eq:delta_target}
\end{equation}
so a model $f_\theta$ produces
\begin{equation}
\widehat{P}_{t+1}^{(i)} = P_t^{(i)} + f_\theta(\mathbf{x}_t^{(i)}).
\label{eq:delta_forecast}
\end{equation}
Equation~\eqref{eq:delta_target} defines the learning target, and Equation~\eqref{eq:delta_forecast} maps the predicted change back to price space. This formulation forces the model to learn an adjustment to the current close rather than relearn the price level. As a diagnostic ablation, we also fit the same models directly to $P_{t+1}^{(i)}$. All evaluation is performed after transforming predictions back to the original CNY price scale.

To illustrate the timing sequence, consider bond 110086 at the start of the 2024H2 test window. At the market close on 1 July 2024 ($P_t=89.670$ CNY), the model generates a forecast for the 2 July close ($P_{t+1}=92.691$ CNY). The persistence benchmark predicts 89.670 CNY (absolute error: 3.021 CNY), whereas the five-seed small MLP ensemble predicts 89.872 CNY (absolute error: 2.819 CNY). Across all experiments, performance evaluation and loss differentials are computed directly on the original CNY price scale.

\subsection{Random Walk and Statistical Baselines}
\label{sec:random_walk}

The primary benchmark is persistence (the random walk),
\begin{equation}
\widehat{P}_{t+1}^{(i),\mathrm{RW}}=P_t^{(i)},
\label{eq:rw}
\end{equation}
which corresponds to an expected zero price change. We additionally benchmark four standard econometric baselines: historical drift (which adds each bond's historical mean training return), a pooled AR(1) model regressing $\Delta P_{t+1}$ on $\Delta P_t$, and regularized linear models (Ridge and Elastic Net) trained on the feature space with hyperparameter penalties optimized strictly on validation data.

\subsection{Identity-Free Neural Regressors}
\label{sec:architecture}

To ensure that panel learning reflects generalizable cross-asset relationships rather than memorized asset-specific fixed effects, our neural architectures deliberately exclude bond identity codes and embeddings. This guarantees that predictions depend solely on observed financial covariates, enabling consistent evaluation across both panel-trained and held-out instruments without requiring ad-hoc embedding imputation.

We use two multilayer perceptrons (MLPs). The small network has hidden widths $[64,32]$ and the large network has widths $[256,256,256,128]$. Each hidden layer applies an affine transformation, GELU activation, and dropout with rate 0.1; a final linear unit produces the scalar target prediction. Both networks minimize mean squared error with AdamW (learning rate $3\times10^{-4}$, weight decay $10^{-4}$), batch size 512, gradient clipping at 1, and cosine learning-rate annealing. Training lasts at most 100 epochs, and the checkpoint with the lowest validation RMSE is retained with patience 15. We repeat every neural specification using seeds 42, 123, 777, 2024, and 2026.

\subsection{Pooled, Single-Bond, and LOBO Estimation}
\label{sec:training}

In pooled estimation, one model is trained on all eligible bond-day observations in the training period. In single-bond estimation, the small MLP is refit separately for each of the 90 test-period bonds. Each fit has its own training and validation rows, feature scaler, and target scaler; architecture and optimization settings otherwise match the pooled small MLP exactly. This fixed-recipe comparison holds architecture constant, but it does not by itself isolate a pooling mechanism because pooled and bond-specific data sets imply different updates per epoch, validation noise, and effective regularization.

We therefore add three controls. First, we compare the pooled and single-bond five-seed ensembles directly on identical bond-day rows. Second, we repeat both estimators under exact per-fit budgets of 50, 100, and 250 optimizer updates, validating every 25 updates and retaining the best validation checkpoint. Equal updates are a per-fitted-model control rather than a total-compute equality: single-bond estimation still fits 90 separate parameter sets. Third, we repeat single-bond training with weight decay 0 and $10^{-3}$ around the default $10^{-4}$. These checks support a forecast-error comparison but do not identify changes in parameter variance.

For leave-one-bond-out (LOBO) evaluation, we select the ten test-period bonds with the fewest training observations. For each fold, the held-out bond is removed from both training and validation, the preprocessing objects and large MLP are refit on the remaining bonds, and evaluation is restricted to the unseen bond. The model never receives a bond-identity variable, so no untrained or imputed embedding is required.

\subsection{Out-of-Sample Performance and Inference}
\label{sec:inference}

For model $m$, the main relative-performance measure is
\begin{equation}
R^2_{\mathrm{OOS},m}
=1-\frac{\sum_{t,i}(P_{t+1}^{(i)}-\widehat{P}_{t+1,m}^{(i)})^2}
{\sum_{t,i}(P_{t+1}^{(i)}-P_t^{(i)})^2}.
\label{eq:oos_r2}
\end{equation}
A positive value means that the model reduces aggregate squared error relative to the random walk. Equation~\eqref{eq:oos_r2} therefore evaluates incremental accuracy rather than ordinary fit to a persistent price level. We also report RMSE, MAE, the RMSE ratio to the random walk, and directional accuracy.

Because bonds observed on the same date share market shocks, bond-day errors are not treated as independent for inference. For each target date $t$, we first calculate the cross-sectional mean loss differential
\begin{equation}
d_t=\frac{1}{N_t}\sum_i\left[e_{\mathrm{RW},i,t}^{2}-e_{m,i,t}^{2}\right].
\label{eq:loss_diff}
\end{equation}
The primary estimand gives every bond-day equal weight, matching Equation~\eqref{eq:oos_r2}:
\begin{equation}
D_{\mathrm{BD}}
=\frac{\sum_t N_t d_t}{\sum_t N_t}.
\label{eq:bond_day_estimand}
\end{equation}
We estimate Equation~\eqref{eq:bond_day_estimand} through an intercept-only weighted regression of $d_t$ with frequency weights $N_t$ and report a two-sided Newey--West/HAC test with five lags. As a distinct sensitivity estimand, we also report the equal-target-date mean
\begin{equation}
D_{\mathrm{date}}=\frac{1}{T}\sum_t d_t,
\label{eq:date_estimand}
\end{equation}
with an unweighted target-date HAC test. The main holdout contains 66--90 bonds per target date, so the two estimands need not be numerically identical.

For ratio uncertainty, the primary bootstrap draws 5,000 circular blocks of ten complete target-date cross sections. Every draw re-sums both models' squared errors and recomputes Equation~\eqref{eq:oos_r2}; it does not average precomputed date-level ratios. We report a two-sided 95\% interval and a one-sided 95\% upper bound on possible improvement. Sensitivities use HAC lags 0, 1, 5, 10, and 20 and both circular and ordinary moving blocks of length 5, 10, and 20. Because these procedures resample saved out-of-sample predictions without refitting the models, their uncertainty is conditional on the fitted forecasts and does not include model-selection or retraining uncertainty.

For neural models, inference uses the arithmetic mean of the five seed-level predictions; seed-level RMSE means and standard deviations remain visible as measures of optimization variation. Direction is evaluated as a three-class outcome (down, unchanged, or up). We report the realized class shares and majority-class accuracy rather than treating 50\% as an automatic chance benchmark.

%% file: sections/04_experimental_setup.tex
\section{Data and Experimental Setup}
\label{sec:experiments}

\subsection{Data and Forecast Timing}
\label{sec:data}

Our empirical analysis utilizes a panel of 73,806 daily observations across 100 Chinese A-share convertible bonds from 15 April 2022 through 31 December 2025, obtained from the Wind Financial Terminal. The sample represents an actively traded cross-section of the Chinese convertible bond market spanning diverse issuing industries—including advanced manufacturing, information technology, materials, and consumer products—and diverse credit quality ratings. Ninety bonds remain actively traded throughout the out-of-sample holdout period, providing 29,514 test evaluations across 367 distinct trading days.

Each observation is indexed by its information date $t$ (\texttt{asof\_date}) and paired with its immediate next observed trading day $t+1$ (\texttt{target\_date}). Across the 903 unique exchange trading dates in the sample period, this construction yields 71,889 forecastable bond-day pairs.

\input{tables/sample_audit}

Table~\ref{tab:sample_audit} details the sample selection steps. To guarantee that performance differences between compact price-lag models and high-dimensional models reflect model capability rather than varying evaluation samples, our primary benchmark utilizes a common complete-case panel requiring full feature availability ($n=65,314$ after purging). In this common panel, 24,983 observations form the training set (2022--2023), 10,817 observations form the validation set (2024H1), and 29,514 observations serve as the primary out-of-sample test holdout (July 2024 through December 2025). As a parallel robustness check, we also evaluate a maximum-availability sample requiring only price lags ($n=69,701$ across all splits, with 29,803 test observations), confirming that feature-completeness screening does not alter any empirical conclusions.

\subsection{Features and Temporal Hygiene}
\label{sec:features}

We examine two feature spaces. The baseline \emph{price-lag} feature set comprises nine fundamental price variables: the current closing price, the prior close, the one-day price change, and cumulative return momentum over 1, 2, 3, 5, 10, and 20 trading days. The \emph{full feature set} expands this space to 39 variables by incorporating rolling realized volatility, moving average spreads, RSI, MACD, Bollinger-band indicators, price range position, turnover volume, lagged pure-bond value, conversion parity value, conversion premium rates, underlying equity returns and volatility, and rolling bond--stock return correlations.

To eliminate look-ahead bias and information leakage across folds, we enforce three strict econometric controls:
\begin{enumerate}[leftmargin=*,itemsep=2pt]
\item \textbf{Target-date alignment and boundary purging:} An observation is assigned to a specific split (training, validation, or testing) if and only if both its information date $t$ and its target date $t+1$ fall strictly within that window. Observations spanning split boundaries (98 rows between training and validation; 90 rows between validation and testing) are purged, preventing the model from training on labels that overlap future periods.
\item \textbf{Point-in-time covariate lag:} All technical indicators and market variables use information available strictly by the market close on date $t$. Underlying stock metrics and Wind-computed valuation fields are lagged by an additional day to prevent contemporaneous look-ahead.
\item \textbf{Fold-local preprocessing:} All feature and target transformations (standardization and scaling) are fitted exclusively on the training fold of each respective window, single bond, or cross-validation fold, completely insulating the test set from global sample statistics.
\end{enumerate}

\subsection{Main Split and Walk-Forward Regimes}
\label{sec:splitting}

The primary chronological split allocates 2022--2023 to model training, 2024H1 to hyperparameter tuning and model selection, and July 2024 through December 2025 to out-of-sample performance evaluation. Model hyperparameters and early stopping criteria are chosen strictly on the validation period and frozen prior to test evaluation.

To assess stability across distinct market environments, we implement a three-window walk-forward design:
\begin{itemize}[leftmargin=*,itemsep=1pt]
\item \textbf{2024H2:} Train 2022--2023, validate 2024H1, test 2024H2 ($n=10{,}948$);
\item \textbf{2025H1:} Train through 2024H1, validate 2024H2, test 2025H1 ($n=9{,}369$);
\item \textbf{2025H2:} Train through 2024, validate 2025H1, test 2025H2 ($n=9{,}033$).
\end{itemize}
These non-overlapping test windows span 29,350 bond-days, allowing direct inspection of whether performance generalizes across changing equity and interest-rate regimes.

\subsection{Model Specifications and Hyperparameter Tuning}
\label{sec:models}

Our baseline suite includes persistence, historical drift, pooled AR(1), Ridge regression, Elastic Net, and two multilayer perceptron architectures (Small MLP and Large MLP). Ridge penalties are tuned across $\{10^{-3},10^{-2},10^{-1},1,10,100\}$; Elastic Net explores penalty parameters $\{10^{-4},10^{-3},10^{-2}\}$ and mixing ratios $\{0.1,0.5\}$ on validation data.

For neural models, each architecture is trained using five fixed random seeds (42, 123, 777, 2024, 2026). The primary objective predicts price change $\Delta P_{t+1}$ using price lags. We systematically ablate these choices by expanding to full features and by contrasting with direct price-level prediction ($P_{t+1}$). All forecasts are evaluated in price space (CNY) against the identical random-walk forecasts.

\subsection{Panel Pooling and Unseen Instrument Generalization}
\label{sec:transfer_setup}

To examine whether panel pooling provides benefits over instrument-specific learning, we evaluate all 90 convertible bonds active in the test period under two conditions: (i) the shared cross-bond pooled MLP, and (ii) separate single-bond MLPs trained exclusively on each individual security. To isolate the pooling effect from compute differences, we also evaluate both specifications under equal optimizer-update budgets (50, 100, and 250 gradient updates per model).

Finally, we assess generalization to unseen instruments via a leave-one-bond-out (LOBO) design on ten test-eligible bonds with low training availability. For each held-out bond, all historical data are excluded from training and validation, preprocessing is refit on the remaining bonds, and out-of-sample predictions are generated without bond identifiers.

\subsection{Evaluation Metrics and Statistical Inference}
\label{sec:metrics}

We report Root Mean Squared Error (RMSE) and Mean Absolute Error (MAE) in CNY. For neural models, seed-level mean and standard deviations capture optimization variation, while formal statistical inference is conducted on the five-seed ensemble prediction for each bond-day to eliminate seed-selection bias.

Relative forecast accuracy is quantified via out-of-sample $R^2$ ($R^2_{\mathrm{OOS}}$). Statistical significance is assessed using target-date clustered Newey--West (HAC) tests weighted by daily cross-sectional sample counts ($N_t$), directly aligning the test statistic with aggregate squared loss. To accommodate non-Gaussian error distributions and temporal persistence, we compute two-sided 95\% confidence intervals and one-sided 95\% upper bounds using circular block bootstrap resampling across complete target-date cross-sections (with block lengths 5, 10, and 20 days). All row-level forecasts and configuration parameters are fully preserved in the accompanying replication package.

%% file: tables/sample_audit.tex
\begin{table}[H]
\centering
\caption{Sample formation. The price-lag complete row is a maximum-availability sensitivity and is not an intermediate step on the common-sample path. Boundary purging requires both the information date and target date to remain inside the same split.}
\label{tab:sample_audit}
\small
\begin{tabular}{lrrl}
\toprule
Stage & Rows & Removed & Role \\
\midrule
Raw workbook & 73,806 & -- & Source export \\
Valid date, bond ID, and close & 71,989 & 1,817 & Common prerequisite \\
Next-date target available & 71,889 & 100 & Forecastable rows \\
Full-feature complete & 65,502 & 6,387 & Common-sample path \\
Common sample after boundary purge & 65,314 & 188 & Main analysis \\
Price-lag complete & 69,889 & 2,000 & Maximum-sample sensitivity \\
Price-lag sample after boundary purge & 69,701 & 188 & Sensitivity analysis \\
\bottomrule
\end{tabular}
\end{table}

%% file: sections/05_results.tex
\section{Results}
\label{sec:results}

\subsection{Main Holdout: No Robust Improvement over Persistence}
\label{sec:results_main}

\input{tables/main_results}

Table~\ref{tab:main_results} reports the central comparison. The random walk achieves RMSE 2.3577 CNY and MAE 1.2129 CNY over 29,514 bond-days. All alternatives have worse aggregate point estimates. Historical drift and AR(1) produce RMSE 2.3610 and 2.3605; their negative $R^2_{\mathrm{OOS}}$ values are statistically distinguishable from zero under the HAC test and their bootstrap intervals are entirely below zero. In other words, even small systematic adjustments to the current close can cause reliably worse forecasts.

The regularized regressions also fail to improve on persistence, although the evidence is less decisive: Ridge and Elastic Net have $R^2_{\mathrm{OOS}}=-0.0056$ and $-0.0044$, while their bond-day-weighted HAC $p$-values are 0.249 and 0.298 and their bootstrap intervals include zero. These estimates support ``no demonstrated improvement,'' not a claim that the models are significantly worse.

The same qualification applies to the neural models. The small price-change MLP has seed-level RMSE $2.3630\pm0.0016$; averaging its five predictions gives ensemble RMSE 2.3619, $R^2_{\mathrm{OOS}}=-0.0035$, bond-day-weighted HAC $p=0.453$, and a 95\% circular block-bootstrap interval $[-0.0127,0.0057]$. The large MLP has seed-level RMSE $2.3664\pm0.0058$ and ensemble RMSE 2.3645, with $R^2_{\mathrm{OOS}}=-0.0057$, $p=0.248$, and interval $[-0.0154,0.0038]$. Equal-target-date HAC $p$-values are similar at 0.418 and 0.224. Adding all features to the large MLP does not change the conclusion. There is therefore no evidence that the pooled MLPs improve expected squared forecast loss relative to persistence.

The one-sided uncertainty bounds quantify the negative result. Conditional on the fitted forecasts, the 95\% upper bounds for $R^2_{\mathrm{OOS}}$ are 0.0045 for the small MLP and 0.0022 for the large MLP. The experiment therefore rules out improvements larger than approximately 0.45\% and 0.22\% of random-walk MSE under its conditional resampling design, while retaining uncertainty about smaller gains. HAC lags from 0 to 20, ordinary versus circular blocks, and block lengths from 5 to 20 produce the same main conclusion; Appendix~\ref{app:inference_sensitivity} reports the sensitivity values.

\subsection{Target Formulation Matters More than Model Capacity}
\label{sec:results_ablation}

\input{tables/target_feature_ablation}

Table~\ref{tab:target_feature_ablation} separates the target and feature choices. All four price-change MLP specifications remain close to the random walk: their seed-level mean RMSE values range from 2.3630 to 2.3683, and every bootstrap interval contains zero. The full feature set worsens the small MLP and modestly improves the large MLP relative to its reduced-feature version, but neither full-feature model beats persistence. These results show that the tested MLP pipelines do not extract stable incremental value from the additional covariates; they do not establish that the variables contain no predictive information under other representations or models.

Directly fitting the next-day price level is substantially less reliable. With price lags, the small and large MLPs have seed-level mean RMSE $2.5461\pm0.0472$ and $2.6799\pm0.1024$. Their ensemble $R^2_{\mathrm{OOS}}$ values are $-0.1042$ and $-0.2171$, and both HAC tests reject equality with the random walk. With full features, the small and large direct-price MLPs deteriorate to ensemble $R^2_{\mathrm{OOS}}=-0.5212$ and $-0.5036$. Under this training setup, residual parameterization around the identity map is therefore much more reliable than direct level fitting. The comparison reflects parameterization, scaling, regularization, and optimization jointly; it does not show that a price-level target is intrinsically invalid.

The maximum-availability price-lag sample does not rescue the neural models. It adds 289 test observations, for a total of 29,803, and changes the random-walk RMSE from 2.3577 to 2.4019 because the added rows are comparatively difficult. The small and large ensembles still have negative $R^2_{\mathrm{OOS}}$ values of $-0.0042$ and $-0.0028$, with HAC $p$-values of 0.362 and 0.490 and bootstrap intervals that cross zero. The main negative conclusion is therefore not created by requiring full-feature completeness.

\subsection{Walk-Forward Stability}
\label{sec:results_wf}

\input{tables/walk_forward}

The non-overlapping walk-forward results in Table~\ref{tab:walkforward} show why a single favorable period would be misleading. In 2024H2, the five-seed ensembles of the small and large MLPs achieve small positive $R^2_{\mathrm{OOS}}$ values of 0.0045 and 0.0048. Neither gain is statistically distinguishable from zero: the HAC $p$-values are 0.414 and 0.324, and both bootstrap intervals cross zero. Both signs reverse in 2025H1 and 2025H2. For the large MLP in 2025H2, the ratio bootstrap interval is entirely negative, although the asymptotic HAC test remains marginal at $p=0.069$. The difference is not a contradiction: the HAC test targets a weighted mean loss differential, while the bootstrap recomputes a nonlinear error ratio under a finite, nonstationary sequence of date blocks.

Aggregating the three windows gives $R^2_{\mathrm{OOS}}=-0.0035$ for the small MLP and $-0.0116$ for the large MLP. Their HAC $p$-values are 0.370 and 0.117, and neither confidence interval supports an improvement. The 2024H2 gains therefore do not generalize across adjacent market periods. Random-walk RMSE rises from approximately 2.00 in the first two windows to 3.01 in 2025H2, and the neural deterioration is strongest in that higher-error regime.

\subsection{Pooling Stabilizes Estimation but Does Not Beat the Benchmark}
\label{sec:results_pooling}

\input{tables/transfer_results}

\begin{figure}[t]
  \centering
  \includegraphics[width=\linewidth]{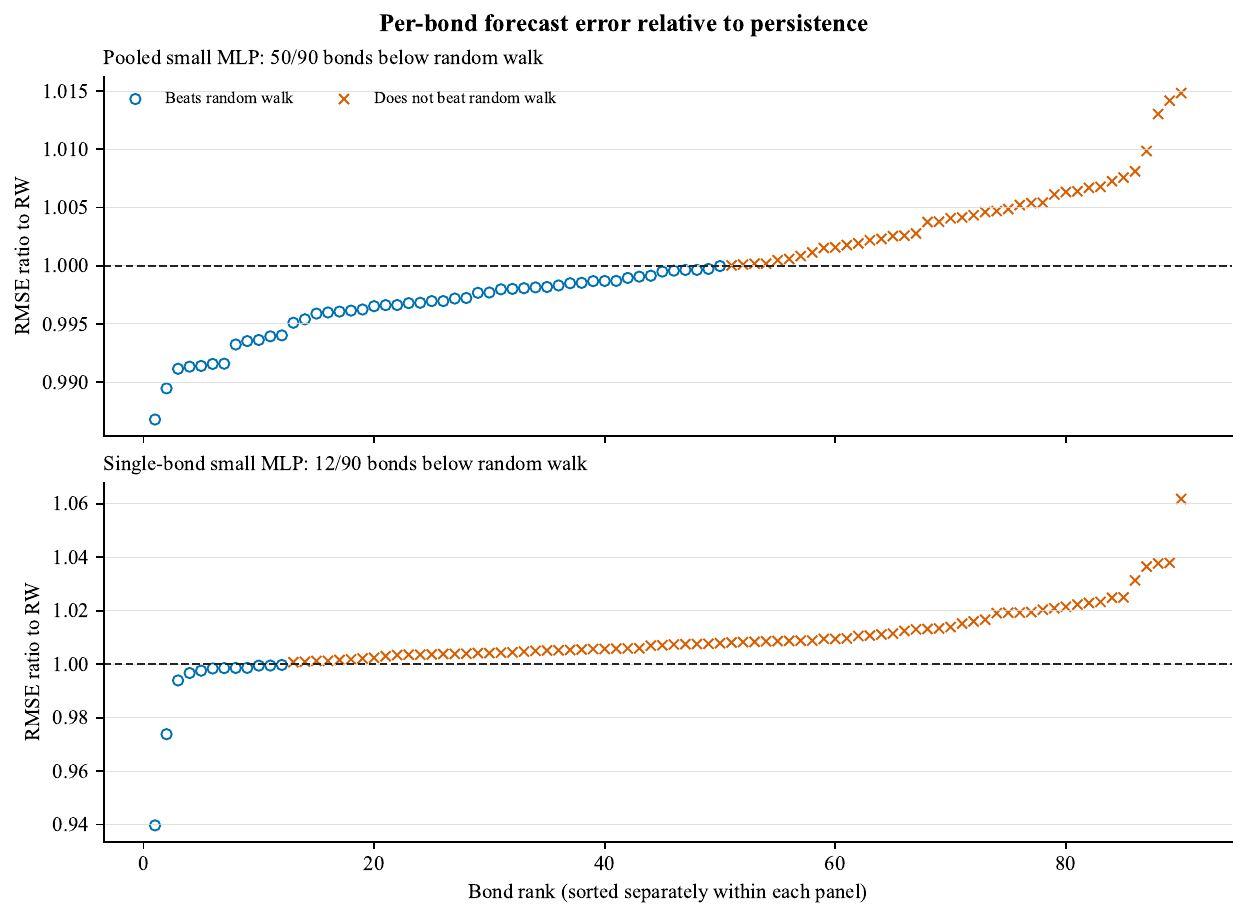}
  \caption{\textbf{Per-bond RMSE relative to the random walk.} Each marker is one bond's mean RMSE across five seeds divided by that bond's random-walk RMSE. Blue open circles denote ratios below one; orange crosses denote ratios at or above one. Bonds are sorted separately within each panel. Pooling moves the distribution toward one and increases the number of bond-level wins, but does not improve aggregate RMSE.}
  \label{fig:per_bond_ratio}
\end{figure}

Table~\ref{tab:transfer_results} and Figure~\ref{fig:per_bond_ratio} distinguish the pooled-versus-single comparison from the harder claim of incremental predictability. Mean seed-level RMSE is 2.3630 for the pooled small MLP and 2.3747 when the same network is fitted separately to each bond. Nevertheless, both remain worse than the aggregate random walk. On a bond-by-bond basis, the pooled model has lower mean RMSE than persistence for 50 of 90 bonds and a median RMSE ratio of 0.9995, whereas single-bond training wins for only 12 bonds and has a median ratio of 1.0072. These descriptive comparisons do not establish a pooling mechanism.

\input{tables/paired_pooling}

Table~\ref{tab:paired_pooling} supplies the direct inferential comparison. Averaging across seeds gives pooled and single-bond RMSE values of 2.3619 and 2.3655 on identical observations, a pooled RMSE advantage of 0.153\%. The corresponding $R^2_{\mathrm{pooled|single}}$ is 0.0031, but the HAC $p$-value is 0.324 and the bootstrap interval $[-0.0026,0.0088]$ crosses zero. Pooling has a favorable point estimate under the default recipe, but its advantage is not statistically resolved.

\input{tables/training_budget}

The default recipe also exposes each pooled model to more updates per fit: the pooled small MLP records 980--1,323 optimizer updates, whereas bond-specific fits record 16--100. Table~\ref{tab:training_budget} therefore compares exact per-fit budgets. Pooling is marginally worse at 50 updates and marginally better at 100 and 250 updates, but every direct interval crosses zero. All six budget-controlled estimators remain worse than the random walk. Moreover, changing single-bond weight decay from 0 through $10^{-3}$ changes ensemble RMSE by less than $10^{-6}$. The evidence is consistent with pooled data limiting some bond-specific overfitting, but it does not identify reduced parameter variance or a robust forecast advantage.

\begin{figure}[t]
  \centering
  \includegraphics[width=\linewidth]{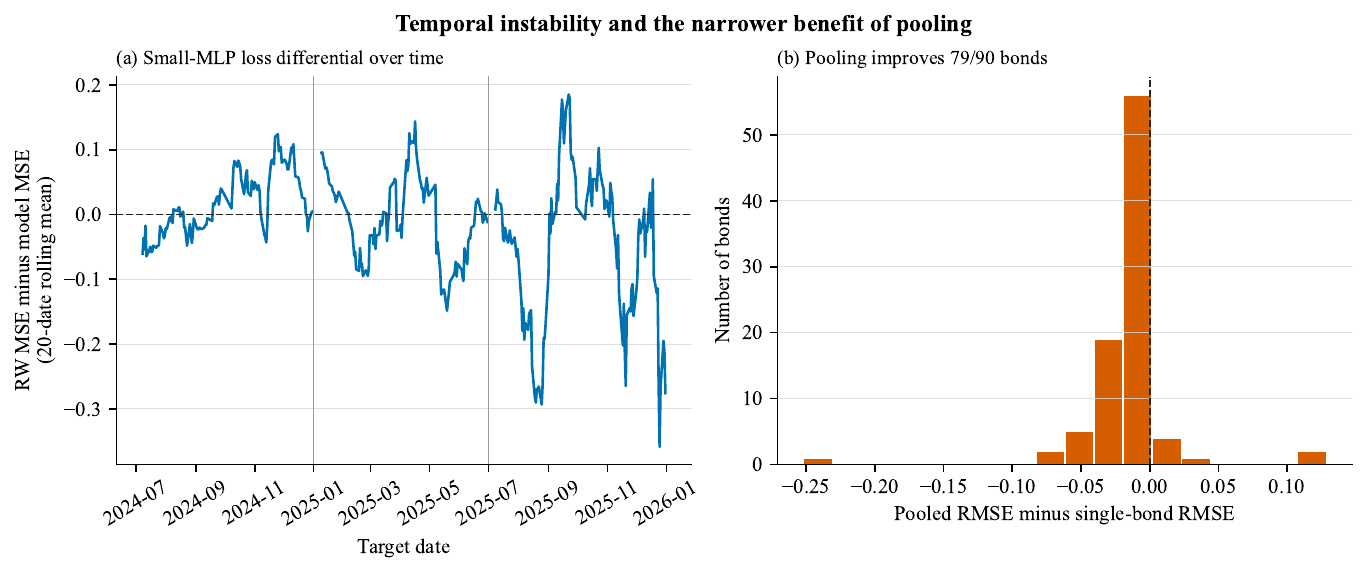}
  \caption{\textbf{The forecast advantage is temporally unstable, whereas pooling usually improves neural estimation.} Panel (a) shows the 20-target-date rolling cross-sectional mean squared-loss differential, random walk minus the five-seed small-MLP ensemble; positive values favor the MLP, and vertical lines separate the half-year test windows. Panel (b) shows the distribution across 90 bonds of pooled-small-MLP RMSE minus single-bond-small-MLP RMSE; negative values favor pooling. Both panels are computed from saved out-of-sample predictions without model refitting.}
  \label{fig:error_diagnostics}
\end{figure}

Figure~\ref{fig:error_diagnostics} shows why the aggregate and pooling conclusions can coexist. Panel (a) plots the 20-target-date rolling mean of the daily squared-loss differential, with positive values favoring the small pooled MLP. The differential changes sign over time rather than remaining persistently positive. Panel (b) compares the pooled and single-bond small MLP bond by bond: pooling lowers mean seed-level RMSE for 79 of 90 bonds, with a median difference of $-0.0102$ CNY, even though the pooled model does not beat the random walk in aggregate.

\input{tables/error_concentration_bonds}

A closer inspection of model loss reveals an extreme concentration of forecast failure across the cross-section. As documented in Table~\ref{tab:error_concentration_bonds}, just five convertible bonds account for 74.4\% (848.99 CNY$^2$) of the total positive excess sum of squared errors generated by the small pooled ensemble relative to persistence across the entire 90-bond test panel. Across all bonds where the model improved upon persistence, cumulative squared error savings totaled 560.20 CNY$^2$. However, this aggregate gain was entirely erased by the massive 1,141.30 CNY$^2$ loss differential accumulated across the underperforming tail, resulting in a net sample-wide disadvantage of 581.20 CNY$^2$.

The underlying financial dynamics of these five securities illuminate why generic neural panel models struggle in this market. In China's convertible bond market, when securities transition into speculative ``double-high'' regimes (characterized by high absolute prices and steep conversion premia), trading behavior decouples from continuous price paths. As shown in Table~\ref{tab:error_concentration_bonds}, bonds 123162, 113657, and 127070 traded across extraordinary ranges during the holdout period—surging from near-par values (84--97 CNY) to speculative peaks of 325--343 CNY—and experienced extreme single-day absolute price jumps between 46.65 and 50.13 CNY. Similarly, bonds 113651 and 123152 exhibited single-day jumps of 27.87 and 25.30 CNY. These discrete price shocks typically coincide with underlying equity speculative frenzies, board announcements regarding conversion price downward resets, or approaching deadlines for conditional mandatory redemption calls. 

Under quadratic loss, a continuous neural regressor that attempts to smooth or extrapolate momentum across these jump dates incurs devastating squared error penalties. While persistence likewise fails to anticipate discrete jumps, it commits no directional overshoot. Consequently, the aggregate failure of cross-bond neural forecasting is not a symptom of broad cross-sectional miscalibration, but rather a structural vulnerability to unmodeled, contract-specific tail events.

\subsection{Generalization to Unseen Instruments (LOBO)}
\label{sec:results_lobo}

For the ten held-out convertible bonds evaluated in the leave-one-bond-out (LOBO) experiment, the large pooled MLP achieves a mean RMSE of $2.0056\pm0.0033$, compared to 2.0181 for persistence. Six of the ten bonds exhibit lower error under panel learning, yielding an aggregate $R^2_{\mathrm{OOS}}$ of 0.0130. While this represents the sole positive aggregate point estimate in our study, the evidence remains exploratory: the primary target-date clustered HAC test yields $p=0.090$, and the 10-day block bootstrap confidence interval spans zero ($[-0.0008, 0.0248]$). Sensitivity checks across block lengths indicate that significance is highly sensitive to the temporal dependence structure. Given the restricted sample size of ten instruments, these findings provide an intriguing hypothesis for cross-sectional knowledge transfer to newly listed securities, but fall short of establishing robust, tradeable out-of-sample predictability.

\subsection{Directional Accuracy}
\label{sec:results_da}

Directional accuracy provides no complementary evidence of useful signal. Up, down, and unchanged moves comprise 52.07\%, 47.63\%, and 0.30\% of the main holdout, so the majority-direction benchmark is 52.07\%. The small and large price-lag ensembles obtain 50.53\% and 49.57\%, while the full-feature ensembles obtain 47.67\% and 47.59\%. The price-lag models also predict declines on 63.69\% and 73.45\% of observations, revealing directional miscalibration rather than a hidden advantage obscured by squared loss.

%% file: tables/main_results.tex
\begin{table}[H]
\centering
\caption{Main common-sample holdout. Neural seed RMSE is mean $\pm$ standard deviation across five fits; ensemble RMSE, MAE, $R^2_{\mathrm{OOS}}$, HAC $p$, and the block-bootstrap interval use the five-seed average prediction. Positive $R^2_{\mathrm{OOS}}$ favors the model over the random walk.}
\label{tab:main_results}
\small
\resizebox{\textwidth}{!}{%
\begin{tabular}{lllrrrrrrrr}
\toprule
Model & Target & Features & $T$ & Seed RMSE & Ensemble RMSE & Ensemble MAE & RMSE/RW & $R^2_{\mathrm{OOS}}$ & HAC $p$ & Bootstrap 95\% CI \\
\midrule
Random walk & price & persistence & 367 & 2.3577 & 2.3577 & 1.2129 & 1.0000 & 0.0000 & -- & -- \\
Historical drift & delta & price lags & 367 & 2.3610 & 2.3610 & 1.2182 & 1.0014 & -0.0027 & $<0.001$ & [-0.0041, -0.0014] \\
AR(1) & delta & price lags & 367 & 2.3605 & 2.3605 & 1.2164 & 1.0012 & -0.0023 & 0.018 & [-0.0042, -0.0006] \\
Ridge & delta & price lags & 367 & 2.3643 & 2.3643 & 1.2154 & 1.0028 & -0.0056 & 0.249 & [-0.0146, 0.0034] \\
Elastic Net & delta & price lags & 367 & 2.3629 & 2.3629 & 1.2145 & 1.0022 & -0.0044 & 0.298 & [-0.0120, 0.0033] \\
Small pooled MLP & delta & price lags & 367 & 2.3630 $\pm$ 0.0016 & 2.3619 & 1.2140 & 1.0018 & -0.0035 & 0.453 & [-0.0127, 0.0057] \\
Large pooled MLP & delta & price lags & 367 & 2.3664 $\pm$ 0.0058 & 2.3645 & 1.2158 & 1.0029 & -0.0057 & 0.248 & [-0.0154, 0.0038] \\
Large pooled MLP & delta & full & 367 & 2.3633 $\pm$ 0.0045 & 2.3627 & 1.2177 & 1.0021 & -0.0042 & 0.244 & [-0.0110, 0.0025] \\
\bottomrule
\end{tabular}%
}
\begin{minipage}{0.98\textwidth}\footnotesize
The primary estimand gives each bond-day equal weight. HAC tests use target-date clusters and $N_t$ frequency weights with five Newey--West lags. Confidence intervals resample complete target-date blocks, re-sum squared errors, and recompute the OOS-$R^2$ ratio in each draw.
\end{minipage}
\end{table}

%% file: tables/target_feature_ablation.tex
\begin{table}[H]
\centering
\caption{Target and feature ablation. Seed RMSE is mean $\pm$ standard deviation across five fits; ensemble metrics and inference use the five-seed average prediction.}
\label{tab:target_feature_ablation}
\small
\resizebox{\textwidth}{!}{%
\begin{tabular}{lllrrrrrrr}
\toprule
Model & Target & Features & $T$ & Seed RMSE & Ensemble RMSE & Ensemble MAE & $R^2_{\mathrm{OOS}}$ & HAC $p$ & Block-bootstrap 95\% CI \\
\midrule
Small pooled MLP & delta & price lags & 367 & 2.3630 $\pm$ 0.0016 & 2.3619 & 1.2140 & -0.0035 & 0.453 & [-0.0127, 0.0057] \\
Large pooled MLP & delta & price lags & 367 & 2.3664 $\pm$ 0.0058 & 2.3645 & 1.2158 & -0.0057 & 0.248 & [-0.0154, 0.0038] \\
Small pooled MLP & delta & full & 367 & 2.3683 $\pm$ 0.0021 & 2.3668 & 1.2226 & -0.0077 & 0.179 & [-0.0183, 0.0030] \\
Large pooled MLP & delta & full & 367 & 2.3633 $\pm$ 0.0045 & 2.3627 & 1.2177 & -0.0042 & 0.244 & [-0.0110, 0.0025] \\
Small pooled MLP & price & price lags & 367 & 2.5461 $\pm$ 0.0472 & 2.4775 & 1.2730 & -0.1042 & $<0.001$ & [-0.1461, -0.0636] \\
Large pooled MLP & price & price lags & 367 & 2.6799 $\pm$ 0.1024 & 2.6011 & 1.3196 & -0.2171 & $<0.001$ & [-0.3003, -0.1440] \\
Small pooled MLP & price & full & 367 & 3.0011 $\pm$ 0.1740 & 2.9079 & 1.5854 & -0.5212 & $<0.001$ & [-0.7114, -0.3670] \\
Large pooled MLP & price & full & 367 & 2.9689 $\pm$ 0.1274 & 2.8911 & 1.5121 & -0.5036 & $<0.001$ & [-0.6854, -0.3470] \\
\bottomrule
\end{tabular}%
}
\end{table}

%% file: tables/walk_forward.tex
\begin{table}[H]
\centering
\caption{Non-overlapping walk-forward tests. RMSE is mean $\pm$ standard deviation across five seeds; inference uses the five-seed average prediction.}
\label{tab:walkforward}
\small
\resizebox{\textwidth}{!}{%
\begin{tabular}{llrrrrr}
\toprule
Window & Model & $T$ & RMSE & $R^2_{\mathrm{OOS}}$ & HAC $p$ & Block-bootstrap 95\% CI \\
\midrule
2024H2 & Random walk & 124 & 2.0001 & 0.0000 & -- & -- \\
2024H2 & Small pooled MLP & 124 & 1.9973 $\pm$ 0.0032 & 0.0045 & 0.414 & [-0.0064, 0.0136] \\
2024H2 & Large pooled MLP & 124 & 1.9968 $\pm$ 0.0027 & 0.0048 & 0.324 & [-0.0035, 0.0127] \\
2025H1 & Random walk & 116 & 2.0060 & 0.0000 & -- & -- \\
2025H1 & Small pooled MLP & 116 & 2.0119 $\pm$ 0.0043 & -0.0035 & 0.613 & [-0.0181, 0.0074] \\
2025H1 & Large pooled MLP & 116 & 2.0115 $\pm$ 0.0024 & -0.0042 & 0.637 & [-0.0234, 0.0110] \\
2025H2 & Random walk & 125 & 3.0068 & 0.0000 & -- & -- \\
2025H2 & Small pooled MLP & 125 & 3.0215 $\pm$ 0.0132 & -0.0078 & 0.209 & [-0.0181, 0.0045] \\
2025H2 & Large pooled MLP & 125 & 3.0488 $\pm$ 0.0150 & -0.0238 & 0.069 & [-0.0456, -0.0017] \\
All windows & Small pooled MLP & 365 & 2.3643 $\pm$ 0.0048 & -0.0035 & 0.370 & [-0.0110, 0.0040] \\
All windows & Large pooled MLP & 365 & 2.3748 $\pm$ 0.0056 & -0.0116 & 0.117 & [-0.0253, 0.0009] \\
\bottomrule
\end{tabular}%
}
\end{table}

%% file: tables/transfer_results.tex
\begin{table}[H]
\centering
\caption{Pooling, bond-specific training, and leave-one-bond-out (LOBO) evaluation. Neural RMSE is mean $\pm$ standard deviation across five seeds. Per-bond wins compare mean seed-level RMSE with each bond's random walk.}
\label{tab:transfer_results}
\small
\resizebox{\textwidth}{!}{%
\begin{tabular}{lrrrrrr}
\toprule
Evaluation & Bonds & RMSE & Median bond ratio & Wins & $R^2_{\mathrm{OOS}}$ & HAC $p$ \\
\midrule
Pooled, observed bonds & 90 & 2.3630 $\pm$ 0.0016 & 0.9995 & 50/90 & -0.0035 & 0.453 \\
Single-bond training & 90 & 2.3747 $\pm$ 0.0036 & 1.0072 & 12/90 & -0.0066 & 0.108 \\
LOBO, unseen bonds & 10 & 2.0056 $\pm$ 0.0033 & 0.9977 & 6/10 & 0.0130 & 0.090 \\
\bottomrule
\end{tabular}%
}
\end{table}

%% file: tables/paired_pooling.tex
\begin{table}[H]
\centering
\caption{Direct paired comparison of pooled and bond-specific small MLPs. Both forecasts are five-seed ensembles evaluated on identical bond-day rows. Positive $R^2$ favors pooling.}
\label{tab:paired_pooling}
\small
\begin{tabular}{rrrrrrr}
\toprule
Bond-days & $T$ & Pooled RMSE & Single RMSE & $R^2_{\mathrm{pooled|single}}$ & HAC $p$ & Bootstrap 95\% CI \\
\midrule
29514 & 367 & 2.3619 & 2.3655 & 0.0031 & 0.324 & [-0.0026, 0.0088] \\
\bottomrule
\end{tabular}
\end{table}

%% file: tables/training_budget.tex
\begin{table}[H]
\centering
\caption{Direct pooled-versus-single comparison under exact per-fit optimizer-update budgets. Positive $R^2_{\mathrm{pooled|single}}$ and positive RMSE improvement favor pooling. Every row uses five-seed ensembles on the same 29,514 bond-days and 367 target dates.}
\label{tab:training_budget}
\small
\resizebox{\textwidth}{!}{%
\begin{tabular}{rrrrrrr}
\toprule
Updates & Pooled RMSE & Single RMSE & Pooled RMSE improvement & $R^2_{\mathrm{pooled|single}}$ & HAC $p$ & Bootstrap 95\% CI \\
\midrule
50  & 2.3621 & 2.3616 & $-0.018\%$ & -0.0004 & 0.911 & [-0.0072, 0.0058] \\
100 & 2.3614 & 2.3659 & 0.190\%    &  0.0038 & 0.382 & [-0.0051, 0.0122] \\
250 & 2.3629 & 2.3694 & 0.276\%    &  0.0055 & 0.202 & [-0.0030, 0.0138] \\
\bottomrule
\end{tabular}%
}
\begin{minipage}{0.98\textwidth}\footnotesize
The budget is equal per fitted model, not in total compute: single-bond estimation fits 90 distinct networks per seed. Validation occurs every 25 updates, and the best validation checkpoint is retained.
\end{minipage}
\end{table}

%% file: tables/error_concentration_bonds.tex
\begin{table}[H]
\centering
\caption{Extreme error concentration in the out-of-sample holdout (July 2024--December 2025). The table lists the five convertible bonds generating the largest positive excess sum of squared errors ($\Delta\text{SSE} = \text{SSE}_{\text{model}} - \text{SSE}_{\text{RW}}$) for the small pooled MLP ensemble relative to persistence. Together, these five instruments account for 74.4\% of all positive excess squared error across the 90-bond holdout panel.}
\label{tab:error_concentration_bonds}
\small
\resizebox{\textwidth}{!}{%
\begin{tabular}{llrrrrrrr}
\toprule
Bond Code & Underlying Stock & Test Obs & Price Range (CNY) & Max 1-Day Jump (CNY) & RW SSE & Model SSE & Excess SSE & Share (\%) \\
\midrule
123162 & 300486.SZ & 278 & [84.15, 343.10] & 50.13 & 8,544.6 & 8,793.4 & +248.82 & 21.8\% \\
113657 & 603601.SH & 367 & [97.48, 325.77] & 48.98 & 7,397.1 & 7,604.6 & +207.54 & 18.2\% \\
127070 & 001203.SZ & 367 & [87.97, 335.00] & 46.65 & 15,067.4 & 15,251.1 & +183.62 & 16.1\% \\
113651 & 603992.SH & 367 & [113.01, 292.51] & 27.87 & 8,271.1 & 8,386.7 & +115.62 & 10.1\% \\
123152 & 300727.SZ & 187 & [103.33, 179.00] & 25.30 & 3,680.3 & 3,773.7 & +93.39 & 8.2\% \\
\midrule
\multicolumn{2}{l}{\textbf{Top 5 Cumulative Total}} & 1,566 & -- & -- & 42,960.5 & 43,809.5 & \textbf{+848.99} & \textbf{74.4\%} \\
\multicolumn{2}{l}{All 90 Test Bonds Net Total} & 29,514 & -- & -- & 164,057.2 & 164,638.4 & +581.20 & -- \\
\bottomrule
\end{tabular}%
}
\begin{minipage}{0.98\textwidth}\footnotesize
\vspace{4pt}
\textit{Notes:} Excess SSE is the difference between the five-seed pooled small MLP ensemble squared error and the random walk squared error. Total positive excess SSE across all bonds with $\text{SSE}_{\text{model}} > \text{SSE}_{\text{RW}}$ is 1,141.3 CNY$^2$, of which 848.99 CNY$^2$ (74.4\%) is generated by these five securities. Price range and maximum 1-day absolute jump are observed during the test holdout.
\end{minipage}
\end{table}

%% file: sections/06_discussion.tex
\section{Discussion and Economic Implications}
\label{sec:discussion}

\subsection{Why Beating Persistence is Formidable: Equity Coupling and Market Efficiency}
\label{sec:discussion_interpretation}

Our central empirical finding—that cross-bond neural networks fail to reliably outperform the persistence benchmark—is deeply grounded in financial economics. Convertible bonds are hybrid contingent claims whose secondary market price formation is inextricably linked to the underlying equity. Over a daily horizon, a substantial share of convertible bond price variation is governed by contemporaneous stock price movements ($\Delta S_{t+1}$). Because daily equity returns closely approximate a martingale difference sequence under weak-form market efficiency~\citep{fama1970}, any model tasked with forecasting next-day convertible bond prices unconditionally using only lagged information ($t$) is essentially burdened with predicting next-day equity direction.

This structural reality explains why reported successes in earlier literature must be interpreted with caution. In studies where same-day equity features or look-ahead signals are inadvertently admitted, or where models are evaluated merely on high price-level $R^2$, neural networks can appear remarkably accurate by mimicking the identity mapping $\widehat{P}_{t+1} \approx P_t$. As our target ablation demonstrates (Table~\ref{tab:target_feature_ablation}), directly training neural networks to reproduce the price level creates massive estimation variance ($R^2_{\mathrm{OOS}} = -0.1042$ to $-0.5212$), whereas formulating the task in price changes ($\Delta P_{t+1}$) forces the estimator to learn true economic adjustments. Once this identity illusion is eliminated, the remaining signal-to-noise ratio in daily technical and valuation indicators is insufficient to beat the random walk.

\subsection{Cross-Bond Pooling as Variance Shrinkage rather than Signal Discovery}
\label{sec:discussion_pooling}

Our results provide a nuanced answer to the role of cross-sectional panel learning in derivative markets. Estimating a single shared network across 100 instruments lowers out-of-sample RMSE for 79 out of 90 individual bonds compared to estimating separate networks per bond (Figure~\ref{fig:error_diagnostics}b), yielding a median bond RMSE ratio of 0.9995.

Economically, cross-bond pooling acts as an effective shrinkage regularizer. In daily financial panels, individual time series are short, noisy, and subject to severe overfitting when modeled in isolation. Parameter sharing across the panel restricts model capacity, regularizing parameter variance and dampening idiosyncratic noise fitting. However, our paired tests and compute-budget controls (Tables~\ref{tab:paired_pooling} and~\ref{tab:training_budget}) reveal that this shrinkage effect does not translate into statistically significant predictability over the random walk ($p=0.324$). Pooling makes neural models substantially less erratic than isolated single-instrument fits, but it cannot extract market-timing signal where none exists.

\subsection{Contractual Optionality and Discrete Jump Regimes}
\label{sec:discussion_events}

The dramatic concentration of model error—wherein five securities generate 74.4\% of all positive excess squared error (Table~\ref{tab:error_concentration_bonds})—uncovers a critical mismatch between standard neural architectures and the institutional mechanics of convertible bonds. In the Chinese market, securities frequently trade into speculative ``double-high'' regimes, characterized by market prices far exceeding par (surging above 300 CNY) and substantial conversion premia decoupled from pure debt or equity values. 

Such instruments are intensely vulnerable to discrete regime shifts driven by contractual clauses. For instance, when an issuer approaches the threshold for triggering mandatory early redemption (typically when the stock trades at 130\% of the conversion price for 15 out of 30 consecutive days), the threat of forced call at face value can precipitate sudden, catastrophic collapse in the conversion premium. Conversely, board announcements proposing downward conversion price resets inject immediate non-linear upward jumps. 

Standard feedforward neural networks trained with mean squared error implicitly presume continuous, smooth functional mappings between lagged market indicators and future prices. When confronted with discrete, clause-induced price shocks of 25 to 50 CNY in a single session, smooth neural functions suffer severe quadratic loss penalties. While persistence likewise fails to forecast jump timing, it avoids aggressive directional overshooting. This finding demonstrates that the primary failure mode of financial deep learning in convertible bonds is event-driven rather than algorithmic.

\subsection{Promising Frontiers for Quantitative Convertible Bond Research}
\label{sec:discussion_frontiers}

These empirical realities highlight several productive directions for quantitative and machine-learning research in convertible bond markets:

\begin{enumerate}[leftmargin=*,itemsep=3pt]
\item \textbf{Event-conditioned modeling:} Rather than attempting unconditional daily price forecasting across all market regimes, future architectures should explicitly condition on contractual milestones. Integrating point-in-time countdown variables for conditional call triggers, put option windows, and reset board meetings can distinguish tranquil regimes from impending jump states.
\item \textbf{Relative pricing and delta-hedged spreads:} Forecasting the absolute price level of a hybrid security inevitably conflates equity market timing with derivative mispricing. A more natural financial target is the relative pricing residual—such as changes in the conversion premium ($\Delta \text{Premium}_{t+1}$) or delta-hedged returns ($\Delta P_{t+1} - \Delta \times \Delta S_{t+1}$). Modeling these spreads isolates contractual valuation dynamics from underlying stock movements.
\item \textbf{Architectural diversity and decision trees:} While our investigation strictly controls network capacity through small and large MLPs to isolate the pooling mechanism, tree-based panel ensembles (such as LightGBM and CatBoost) and temporal attention mechanisms (Transformers) offer alternative inductive biases. Examining whether tree ensembles, which naturally partition feature space via orthogonal splits, better handle regime switches in convertible bond panels represents a valuable empirical extension.
\end{enumerate}

In summary, out-of-sample persistence should be embraced as the mandatory benchmark for any quantitative model purporting to forecast convertible bond prices. High in-sample explanatory power and localized validation gains do not constitute evidence of tradeable predictive skill. Future progress depends upon aligning statistical architectures with the discrete contractual dynamics that govern hybrid securities.

%% file: sections/07_conclusion.tex
\section{Conclusion}
\label{sec:conclusion}

This study provides a rigorous empirical evaluation of whether cross-bond panel learning can forecast next-day Chinese convertible bond prices more accurately than the random walk benchmark. Utilizing a leak-resistant framework with strict target-date alignment, split boundary purging, fold-local preprocessing, multi-seed neural ensembles, and cross-sectionally clustered HAC inference, we demonstrate that neither feedforward neural networks nor regularized linear models deliver reliable predictive gains over persistence. On the primary 29,514-observation holdout, the random walk achieves an RMSE of 2.3577 CNY, outperforming five-seed small and large price-change MLP ensembles (RMSEs 2.3619 and 2.3645 CNY; out-of-sample $R^2$ of $-0.0035$ and $-0.0057$). Directly predicting price levels rather than changes degrades forecast accuracy substantially by attempting to fit an identity mapping.

Our findings illuminate two vital economic mechanisms governing financial panel learning. First, cross-bond pooling exerts a beneficial variance-stabilizing shrinkage effect over isolated single-bond estimation—lowering test RMSE for 79 out of 90 individual securities—though this stabilization remains insufficient to overcome the near-martingale behavior of daily asset prices. Second, forecast failure is heavily concentrated in the tails: five highly volatile, clause-sensitive convertible bonds account for 74.4\% of the aggregate excess squared error, driven by discrete price jumps during speculative runs into high-convexity regimes. 

These results underscore that high in-sample explanatory power or elevated level $R^2$ cannot substitute for genuine out-of-sample forecasting tests against persistence. For quantitative researchers and practitioners in hybrid security markets, future advances require moving beyond unconditional daily price forecasting toward explicit modeling of discrete contractual events, countdowns to clause triggers, and delta-hedged relative valuation spreads. Persistence should remain the essential starting benchmark against which all future predictive claims are judged.

%% file: sections/08_appendix.tex
\appendix
\section{Inference Sensitivity}
\label{app:inference_sensitivity}

\input{tables/inference_sensitivity}

The main result is invariant to the weighting and dependence settings shown in Table~\ref{tab:inference_sensitivity}. For both MLP sizes, every HAC specification fails to reject equal expected loss and every circular or ordinary moving-block interval contains zero. The similarity between the bond-day and equal-date tests also shows that the unbalanced number of observed bonds per date does not drive the qualitative conclusion, even though the bond-day-weighted test remains the estimand aligned with aggregate $R^2_{\mathrm{OOS}}$.

%% file: tables/inference_sensitivity.tex
\begin{table}[H]
\centering
\caption{Inference sensitivity for the main price-change MLP ensembles. Panel A varies HAC lags under the bond-day-weighted and equal-target-date estimands. Panel B varies block method and length; entries are 95\% $R^2_{\mathrm{OOS}}$ intervals.}
\label{tab:inference_sensitivity}
\small

\textbf{Panel A: HAC $p$-values}\par\smallskip
\begin{tabular}{llrrrrr}
\toprule
Model & Estimand & Lag 0 & Lag 1 & Lag 5 & Lag 10 & Lag 20 \\
\midrule
Small MLP & Bond-day weighted & 0.418 & 0.437 & 0.453 & 0.465 & 0.441 \\
Small MLP & Equal target date & 0.389 & 0.408 & 0.418 & 0.429 & 0.393 \\
Large MLP & Bond-day weighted & 0.233 & 0.242 & 0.248 & 0.262 & 0.237 \\
Large MLP & Equal target date & 0.210 & 0.220 & 0.224 & 0.236 & 0.203 \\
\bottomrule
\end{tabular}

\vspace{8pt}
\textbf{Panel B: target-date block-bootstrap intervals}\par\smallskip
\resizebox{\textwidth}{!}{%
\begin{tabular}{llrrr}
\toprule
Model & Block method & Length 5 & Length 10 & Length 20 \\
\midrule
Small MLP & Circular & [-0.0127, 0.0055] & [-0.0126, 0.0055] & [-0.0122, 0.0055] \\
Small MLP & Ordinary moving & [-0.0131, 0.0048] & [-0.0126, 0.0057] & [-0.0113, 0.0063] \\
Large MLP & Circular & [-0.0153, 0.0035] & [-0.0152, 0.0039] & [-0.0147, 0.0033] \\
Large MLP & Ordinary moving & [-0.0157, 0.0027] & [-0.0154, 0.0037] & [-0.0135, 0.0040] \\
\bottomrule
\end{tabular}%
}
\end{table}